\input{aipcheck}

\documentclass[
    ,final            
  ]
  {aipproc}

\layoutstyle{6x9}

\begin{document}

\title{The End of Amnesia: Measuring the Metallicities of Type Ia SN Progenitors with Manganese Lines in Supernova Remnants}

\classification{26.30.k 95.85.Nv 97.60.Bw}
\keywords      {stars:binaries:close --- supernova remnants --- supernovae:general --- X-rays:ISM}

\author{Carles Badenes}{
  address={Department of Astrophysical Sciences, Princeton University, Peyton Hall, Ivy Lane, Princeton NJ 08544-1001}
  ,altaddress={\textit{Chandra} Fellow}
}

\author{Eduardo Bravo}{
  address={Departament de F\'{i}sica i Enginyeria Nuclear, Universitat Polit\`{e}cnica de Catalunya, Diagonal 647,
  Barcelona 08028, Spain; and Institut d'Estudis Espacials de Catalunya, Campus UAB, Facultat de Ci\`{e}ncies,
  Bellaterra, Barcelona 08193, Spain}
}

\author{John P. Hughes}{
  address={Department of Physics and Astronomy, Rutgers University. 136 Frelinghuysen Rd., Piscataway NJ 08854-8019}
}

\begin{abstract}
  The Mn to Cr mass ratio in supernova ejecta has recently been proposed as a tracer of Type Ia SN progenitor
  metallicity. We review the advantages and problems of this observable quantity, and discuss them in the framework of
  the Tycho Supernova Remnant. The fluxes of the Mn and Cr K$\alpha$ lines in the X-ray spectra of Tycho observed by the
  \textit{Suzaku} satellite suggests a progenitors of supersolar metallicity.
\end{abstract}

\maketitle


\section{Mn/Cr as a Tracer of Metallicity in Type Ia SN Progenitors}

The recent detection of the Cr and Mn K$\alpha$ lines in the X-ray spectrum of the Tycho Supernova Remnant (SNR) by the
Japanese satellite \textit{Suzaku} \cite{tamagawa08:tycho} has opened the possibility to study secondary Fe-peak nuclei
in the shocked ejecta of Type Ia SNRs. The presence of Mn in particular, which has an odd
atomic number, has important implications for constraining several properties of Type Ia progenitors, as described by
Badenes et al. in \cite{badenes08:mntocr}. Under the appropriate conditions, the Mn to Cr mass ratio in the ejecta of
Type Ia SNe ($M_{Mn}/M_{Cr}$) can be used to calculate the metallicity of the SN progenitor. In these Proceedings, we
expand on the information presented in \cite{badenes08:mntocr}, providing more details on the models, the advantages and
limitations of $M_{Mn}/M_{Cr}$ as a tracer of progenitor metallicity, and the applicability of the method to
current and future X-ray observations of SNRs.

The rationale for the use of $M_{Mn}/M_{Cr}$ as a tracer of progenitor metallicity $Z$ was explained in detail in
\cite{badenes08:mntocr} \footnote{Here and in \cite{badenes08:mntocr}, we define $Z$ as the mass fraction of all
  elements heavier than He. Note that the correspondence between this theoretical quantity and a given observational
  tracer of metallicity (like $\mathrm{[Fe/H]}$, for instance) is not necessarily trivial.}. It is based on an argument
by Timmes et al. \cite{timmes03:variations_peak_luminosity_SNIa} that connects the amount of C, N, and O in the
progenitor star (and hence its $Z$) to the trace amount of $^{22}$Ne that is present in the CO white dwarf (WD) that
will eventually explode as a SN Ia. During the explosion itself, both $^{55}$Co and $^{52}$Fe (the parent nuclei of
$^{55}$Mn and $^{52}$Cr, respectively) are synthesized in the incomplete Si burning regime. These nuclides belong to a
quasi-statistical equilibrium group dominated by $^{56}$Ni. While $^{52}$Fe is linked to $^{56}$Ni by the reaction
$\mathrm{^{52}Fe(\alpha,\gamma)^{56}Ni}$, which is not sensitive to $\eta$, $^{55}$Co is linked to $^{56}$Ni by the
$\mathrm{^{55}Co(p,\gamma)^{56}Ni}$ reaction. In this last reaction, the proton abundance is strongly dependent on
$\eta$, with larger values of $\eta$ leading to lower proton abundances, which favors the synthesis of $^{55}$Co in
progenitors with high metallicity. In this context, it is worth mentioning the work of \cite{cescutti08:Mn_evolution},
who find evidence for a metallicity dependent yield of Mn in SN Ia, with Mn synthesis appearing enhanced in high
metallicity stellar systems like the Galactic bulge.

In \cite{badenes08:mntocr}, the nucleosynthetic output of 36 Type Ia SN models calculated with different trace amounts
of $^{22}$Ne in the CO WD was examined to quantify the relationship between $Z$ and $M_{Mn}/M_{Cr}$ (Figure
\ref{fig1}). In these calculations, the inner 0.2 $\mathrm{M_{\odot}}$ of ejecta were not included in the final Mn/Cr
ratio. Inside this region, neutron-rich nuclear statistical equilibrium (NSE) takes place, and minor quantities of Mn
and Cr are produced at a mass ratio that is independent of the value of $Z$. Removal of the n-rich NSE products is
justified by the fact that the reverse shock in most ejecta-dominated Type Ia SNRs in our Galaxy has not reached the
inner 0.2 $\mathrm{M_{\odot}}$ of ejecta, and this material does not appear to mix into the outer layers either during
the SN phase \cite{mazzali07:zorro} or the SNR phase \cite{fesen07:SN1885}. Another argument that supports the exclusion
of the n-rich NSE products from an observational point of view is the absence of the K$\alpha$ line from Ni in the same
exposure of the Tycho SNR that revealed the Mn and Cr lines \cite{tamagawa08:tycho}. If a large amount of n-rich NSE
material (or for that matter, of any kind of NSE material) had been thermalized by the reverse shock, this line would
show up at 7.5 keV in the \textit{Suzaku} spectrum. Fitting a power law to the points shown in Figure \ref{fig1} yields
the relation $M_{Mn}/M_{Cr}=5.3 \times Z^{0.65}$ \cite{badenes08:mntocr}, which is virtually independent of the details
of the explosion dynamics.

\section{The Impact of C simmering on the Mn/Cr ratio}

Before a slowly accreting WD explodes as a SN Ia, there is a $\sim$ 1000 yr long phase of slow C fusion in its core. The
energy input from this `simmering' creates a more or less extended convective region inside the WD. It was pointed out
by \cite{piro08:neutronization_SNIasimmering} that the weak interactions which take place during this simmering phase can
increase the value of $\eta$ in the WD material, albeit only by $\Delta \eta=0.0015$
\cite{chamulak08:reduction_electron_simmering_SNIa}. The impact that this increase of $\eta$ will have on the
$M_{Mn}/M_{Cr}$ ratio will depend on the extent of the overlap between the convective core and the explosive Si burning
region of the ejecta where Mn and Cr are synthesized. Unfortunately, the extent of the convective core in a
pre-explosion CO WD is not known, and cannot be calculated self-consistently without sophisticated simulations
\cite{piro08:neutronization_SNIasimmering2}.

To estimate this impact, we consider two limiting cases. If the entire WD is convective, all the explosive Si burning
products are affected by simmering, and the $M_{Mn}/M_{Cr}$ ratio has a lower bound of 0.4, which does not allow to
measure metallicities below solar (see Figure \ref{fig1}). Another possibility is that the size of the convective core
is limited by the Ledoux criterion to the central C-depleted region created during hydrostatic He-burning, as proposed
by \cite{hoeflich02:runaway}. There are a number of reasons why this seems plausible (see \cite{badenes08:mntocr}), but
there is no way to prove that it is indeed the case. In this scenario, the impact of simmering is only large for
subluminous Type Ia SNe, whose Si-rich regions reach deeper into the SN ejecta. The effect is also stronger at lower
$Z$, because metal-poor stars have larger C-depleted cores \cite{dominguez01:SNIa_progenitors}. Two examples of
simmering-modified models with limited convection are shown in Figure \ref{fig1}. Under these conditions, the impact of
C simmering should be of no concern for the Tycho SNR, because we know both from the historical light curve
\cite{ruiz-lapuente04:TychoSN} and the X-ray emission of the SNR \cite{badenes06:tycho} that the SN of 1572 was not
subluminous, but rather normal or slightly overluminous.

\section{Observations: Tycho and Beyond}

\begin{figure}
  \includegraphics[height=.3\textheight,angle=90]{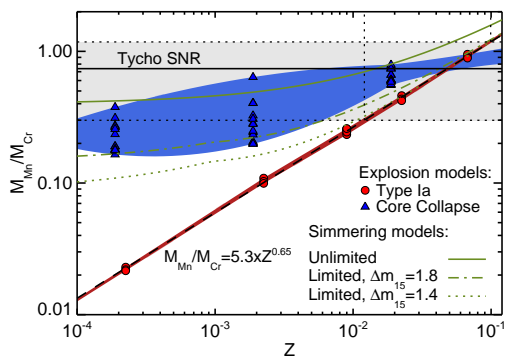}
  \caption{The $M_{Mn}/M_{Cr}$ ratio vs. $Z$ in Type Ia (red circles) and core collapse (blue triangles) SN models
    \cite{badenes08:mntocr,woosley95:core_collapse_models}, together with the observed $M_{Mn}/M_{Cr}$ for Tycho,
    adapted from \cite{badenes08:mntocr}). Some simple models for the impact of simmering are also shown. The solid
    green plot represents the most pessimistic case of a fully convective WD. The dash-dotted and dotted green plots
    represent the more optimistic case of a convective core limited to the C-depleted region of the WD, for a
    subluminous ($\Delta m_{15}=1.8$), and a normal, but faint ($\Delta m_{15}=1.4$) Type Ia SN. In the models with
    limited convection, it has been assumed that the WD progenitor had a ZAMS mass of
    $5\,\mathrm{M_{\odot}}$. \label{fig1}}
\end{figure}

Assuming no impact from C simmering, the metallicity of the Tycho SN progenitor would be $Z=0.048^{+0.051}_{-0.036}$
(from \cite{badenes08:mntocr}, see Figure \ref{fig1}). The $M_{Mn}/M_{Cr}$ ratio is too large to be explained by C
simmering alone, which suggests a solar or supersolar metallicity for the progenitor, regardless of the size of the
convective core. The large uncertainties in the measured values of $M_{Mn}/M_{Cr}$ make it very difficult to completely
discard \textit{some} impact from simmering, specially in the most pessimistic case of unlimited convection. This
underlines the importance of improving the $M_{Mn}/M_{Cr}$ measurements, both by using adequate atomic data for Mn and
Cr (in \cite{badenes08:mntocr}, the specific emissivities had to be interpolated, which introduced a large uncertainty
in the measurement) and by calculating line fluxes from deeper observations that can reduce the statistical
errors. Determining accurate fluxes for such weak lines represents a challenge for the present generation of X-ray
telescopes. An improvement on the measurements presented here is certainly possible for bright SNRs, but more distant
objects will require either much larger collecting areas or a new type of detectors, such as the microcalorimeters being
planned for the \textit{NeXT/ASTRO-H} mission \cite{fujimoto02:calorimeter}. Another exciting prospect for the future is
the extension of this method to other Type Ia SNRs. \textit{Suzaku} has recently detected lines from Mn and Cr in the
spectrum of the Kepler SNR (S. Park, private communication), and the prospects for finding at least Cr in the Galactic
SNR G337.2$-$0.7 are good \cite{rakowski05:G337}, so more measurements will be available soon.


\section{Conclusions}

In \cite{badenes08:mntocr}, the $M_{Mn}/M_{Cr}$ ratio was introduced as a tracer of SN Ia progenitor metallicity. It
might be more appropriate to say that it is a tracer of \textit{neutron excess} in the explosive Si burning region of
Type Ia SNe. This is interesting in its own right, because it also opens a window into the extent of the convective core
of pre-explosion WDs. We hope that more and better observations will let us disentangle the contributions of metallicity
and C simmering to the neutron excess in Type Ia SNe, and that this can help us to understand the lingering mystery of
Type Ia progenitors.


\begin{theacknowledgments}
  Support for this work was provided by NASA through Chandra Postdoctoral Fellowship Award Number
  PF6-70046 issued by the Chandra X-ray Observatory Center, which is operated by the Smithsonian Astrophysical
  Observatory for and on behalf of NASA under contract NAS8-03060. EB is supported by grants AYA2007-66256 and
  AYA2005-08013-C03-01. JPH is partially supported by NASA grant NNG05GP87G.
\end{theacknowledgments}



\bibliographystyle{aipproc}   


\IfFileExists{\jobname.bbl}{}
 {\typeout{}
  \typeout{******************************************}
  \typeout{** Please run "bibtex \jobname" to optain}
  \typeout{** the bibliography and then re-run LaTeX}
  \typeout{** twice to fix the references!}
  \typeout{******************************************}
  \typeout{}
 }

\end{document}